# Manipulating exciton fine-structure in quantum dots with a lateral electric field

B. D. Gerardot, S. Seidl, P. A. Dalgarno, and R. J. Warburton
*School of Engineering and Physical Sciences, Heriot-Watt University, Edinburgh EH14 4AS, UK*

D. Granados and J. M. Garcia,
*Instituto de Microelectronica de Madrid, CNM (CSIC), Isaac Newton 8, PTM, 28760 Tres Cantos, Madrid, Spain*

K. Kowalik and O. Krebs
*CNRS-Laboratoire de Photonique et Nanostructures, Route de Nozay, 91460 Marcoussis, France*

K. Karrai
*Center for NanoScience and Department für Physik der LMU, Geschwister-Scholl-Platz 1, 80539 Munich, Germany*

A. Badolato and P. M. Petroff,
*Materials Department, University of California, Santa Barbara, California 93106*

The fine structure of the neutral exciton in a single self assembled InGaAs quantum dot is investigated under the effect of a lateral electric field. Stark shifts up to 1.5 meV, an increase in linewidth, and a decrease in photoluminescence intensity were observed due to the electric field. We show that the lateral electric field strongly affects the exciton fine structure splitting due to active manipulation of the single particle wave-functions. Remarkably, the splitting can be tuned over large values and through zero.

There is currently great interest in producing entangled photons on demand for applications in quantum information processing [1]. One proposal which has spurred much research is to use the radiative biexciton-exciton cascade in semiconductor quantum dots (QDs) to produce pairs of polarization entangled photons [2]. In an idealized QD, the bright exciton states (M = ± 1) are degenerate. In this case the two decay paths from the biexciton to the vacuum state via the intermediate single exciton are indistinguishable in energy, thus the photons emitted in the radiative cascade are polarization entangled. However, in practice the rotational symmetry of a self-assembled QD is broken and the electron-hole exchange interaction mixes the bright exciton states into a non-degenerate doublet (referred to as a fine-structure splitting, or FSS). This leads to an energetically distinguishable recombination path for the biexciton-exciton cascade. While polarization correlations are observed in the linear basis, the polarization entanglement is destroyed due to the FSS [3, 4]. For photons to be polarization entangled using this scheme, the requirement that the FSS be less than the homogeneous linewidth must be met. The FSS is typically 10-100 μeV, while the homogeneous linewidth of self-assembled InGaAs QDs is ~ 1 μeV [5].

Several techniques have been used to actively manipulate the value of the FSS, including: an in-plane electric [6] or magnetic [7] field, and an in-situ uniaxial stress [8]. Also, it has been reported that QDs which are smaller due to the growth process [9] or subsequent annealing [10] have a smaller FSS. Unfortunately, these QDs are higher in energy and the photons produced by the QD become difficult to distinguish from those produced in the wetting layer. Recently, polarization entangled photons from specific QDs with energetically overlapping bright exciton states have been reported [11, 12]. However, a robust approach that would allow one to actively tune the FSS from a large value to zero for each QD is still necessary in order to realize an event ready entangled photon pair source. To this end we further explore the effect of a lateral electric field on the fine-structure splitting.

There are three basic characteristics of an exciton in a lateral electric field attributed to the quantum confined Stark effect, as has been investigated for quantum wells [13] and QDs [14]: a red-shift in recombination energy, a decreased oscillator strength, and a decreased exciton lifetime due to an increase in non-radiative carrier tunnelling probability. Additionally, electric fields also affect the FSS of a self-assembled QD [6], and while a thorough understanding is still lacking the expectation of such an effect is motivated by the following. The long-range component of the exchange interaction greatly affects the FSS, and is largely determined by the electron-hole overlap [15]. Self-assembled QDs which are elongated along the [1-10] direction impart an asymmetry onto the electron and hole wave-functions. Therefore, an in-plane electric field changes the single particle wave-functions thus altering the electron-hole overlap and affecting the FSS. While previous results demonstrated a small change in the FSS [6], we show here a much stronger effect such that the FSS can be tuned over a large range (~ 100 μeV) and the degeneracy of the bright exciton states can be restored.

Our samples consist of a single layer of self-assembled InAs/GaAs QDs grown by molecular beam epitaxy. The ground state exciton energies of the QDs are inhomogeneously broadened around 950 nm. Two CrNiAu gate electrodes, each 200 μm wide and separated by a 15 μm channel, were deposited using conventional photolithography. The gate fingers were aligned parallel to the [1-10] direction of the GaAs wafer. The Schottky contacts are used to induce an electric field in the plane of the QDs, which are located 130 nm below the surface. However, due to the roughness of the gate fingers (~ 3 μm), the exact direction of the electric field relative to the crystallographic axis for each QD is not precisely known. Photoluminescence (PL) was performed using non-resonant excitation with an 850 nm laser diode and a 0.5 m focal length spectrometer equipped with a Si-CCD. The spectral resolution was measured to be ~ 33 μeV (FWHM) using a narrowband laser. The FWHM diameter of the diffraction-limited laser spot on the sample surface is ~ 400 nm due to the use of a hemispherical glass ($n$ = 2.0) solid immersion lens. By measuring the polarization dependence of the PL, we can identify neutral excitons in the QD emission through observation of the FSS. The doublet of orthogonally linear-polarized lines can be resolved with an accuracy of ~ 5 μeV by using a Lorentzian curve-fitting procedure [6, 7]. The accuracy of this technique was verified by comparing the doublet in PL to a measurement on the same QD of the FSS made using transmission spectroscopy (Fig. 1), which has sub-μeV resolution [5]. This comparison was made on a sample which allowed for a vertical electric field as presently required in our transmission experiment.

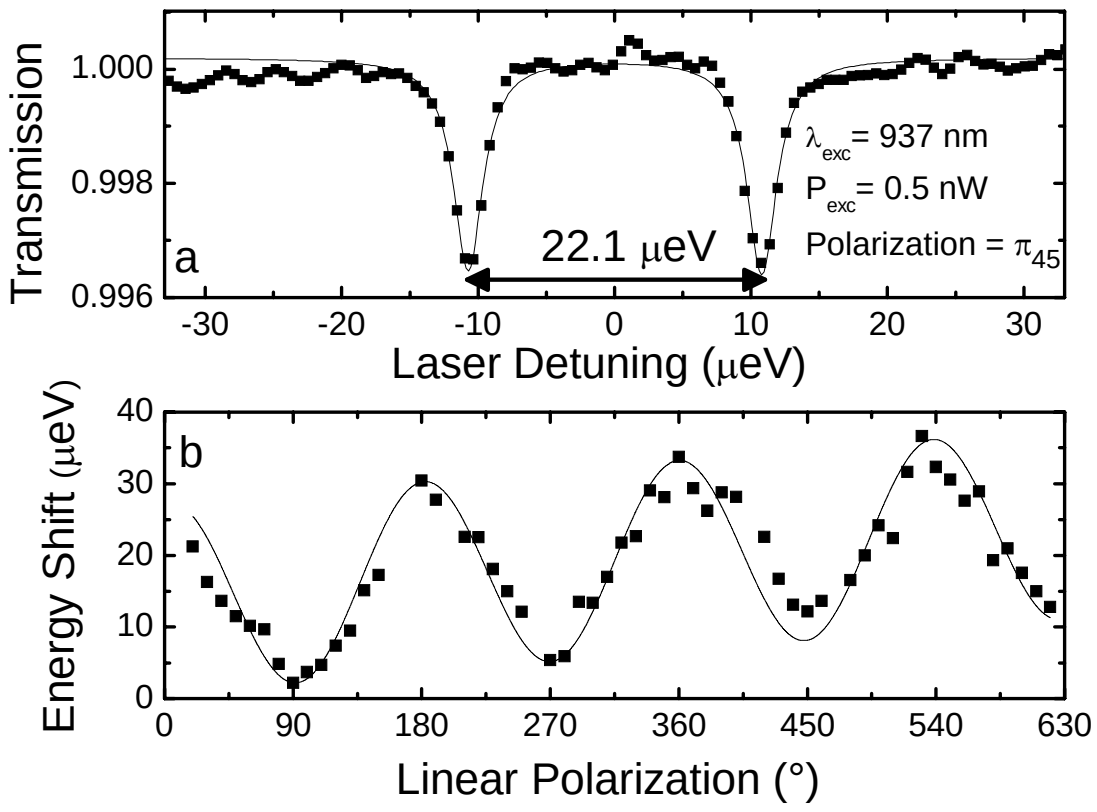

FIG. 1: The fine-structure splitting of the neutral exciton for the same quantum dot measured using (a) laser spectroscopy and (b) polarization dependent photoluminescence. For the laser spectroscopy, the splitting of the two peaks fit with Lorentzian lines is 22.1 μeV. The PL spectra were fit with a Lorentian curve at each polarization. The solid line fit for the energy shift due to polarization is a sin curve with an amplitude (i.e. FSS) of 26.6 ± 3.4 μeV. Additionally, there is a small drift in the PL energy likely caused by a small temperature change in the spectrometer.

The PL from several QDs at different spatial positions in the channel between the two contacts was measured as a function of applied bias. In contrast to previous reports [14], for our device QDs in the centre of the channel show no dependence on the applied bias, signifying that the voltage drop entirely occurs in the depletion region near the gates. By placing a detector below the sample, we can measure the incident laser power in transmission. The transmission power in this setup decreases as the laser spot moves on top of the opaque Au contact. This allows us to measure the ratio of the laser spot on top of the Schottky gate, providing a measurement of the distance of the laser spot center from the gate. In this way, we estimate the lateral depletion width to be ~ 500 nm for the device. Fig. 2 shows the PL from a neutral exciton as a function of applied bias for a QD located ~ 200 nm from the gate. Several features are observed in the spectra. The first is that the PL of the neutral exciton quenches abruptly at ± 1 V which we interpret as a charging event. PL lines shifted in energy appear in the spectra between ± 1 V (not shown), but because of the relatively high QD density in the sample it is difficult to assign different lines to the same dot unambiguously.

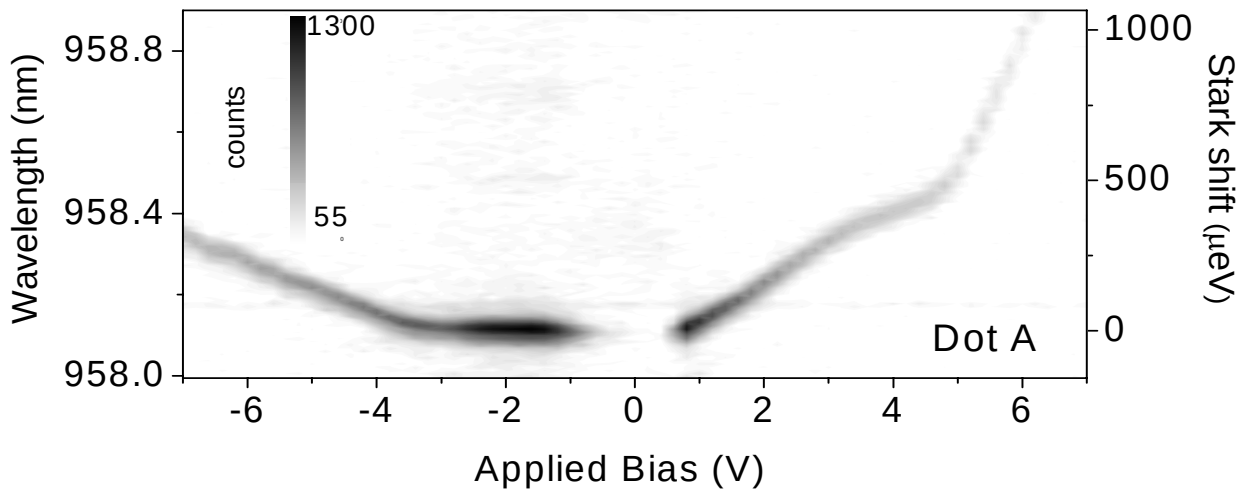

FIG. 2: Low temperature (4 K) photoluminescence of a neutral exciton as a function of applied bias from a QD located in the depletion region of the device. The PL is quenched from ~ +1.0 to -1.0 V due to a charging event, while at larger biases the exciton energy changes due to the Stark effect. In addition to the Stark effect, the PL intensity decreases with increasing bias due to a combination of reduced oscillator strength and carrier tunnelling.

The second feature of note is the large shift in energy, ~ 1 meV for this QD, with applied bias. A dipole in an electric field displays a Stark shift with a quadratic field dependence: $E = E_0 - pF + \beta F^2$, where $E$ is the exciton energy, $F$ the electric field, $p$ the permanent dipole moment, and $\beta$ the polarizability. In our device the electric field strength is highly dependent on QD position and laser power (see below), so essentially we use the QD itself as a local *in situ* probe of the electric field. To estimate $F$ based on the Stark shift observed by PL, we assume a rotationally symmetric (in-plane) parabolic confinement potential. This leads to zero dipole moment in the x-y plane, $p_{xy}$ = 0. We have characterized similar QDs in other samples under the influence of vertical electric fields and find the average lateral extent of the electron ($L_e^{xy}$) and hole ($L_h^{xy}$) wave-function to be 5 and 3 nm, respectively [16]. Using these values, we estimate $\beta$ = -4 μeV/(kV/cm)$^2$ and for a Stark shift of 1 meV, $F$ = 15kV/cm.

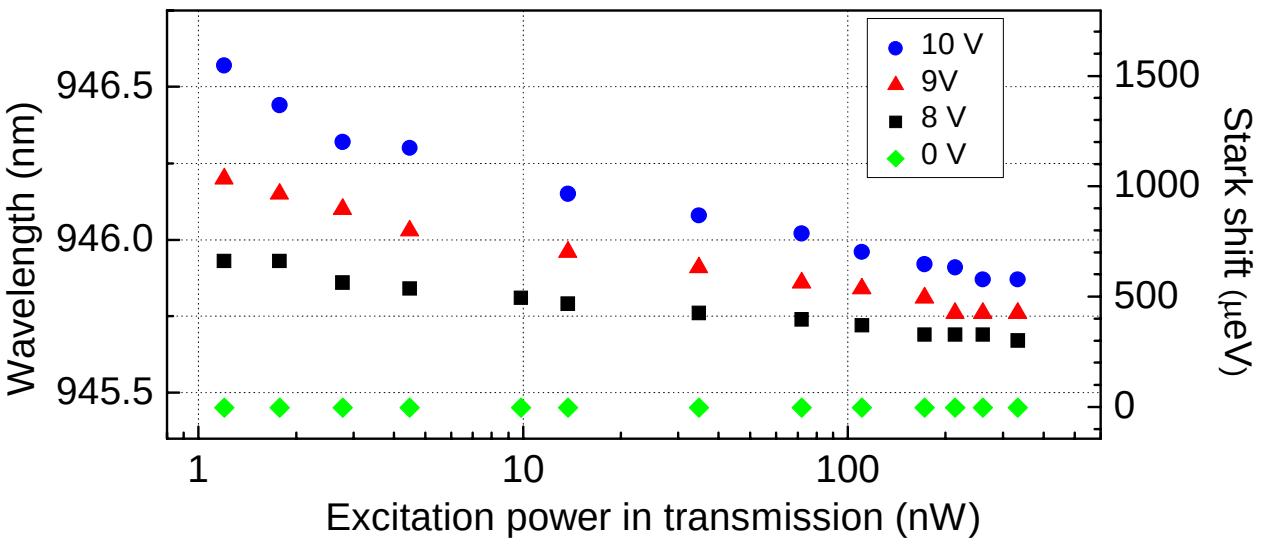

FIG. 3: The effect of carrier screening in the device for different applied biases. At high excitation powers the Stark effect is strongly reduced due to space-charge accumulation in the device. At low powers Stark shifts of ~ 1.5 meV are observed for an applied bias of 10V.

Fig. 3 displays the Stark shift measured as a function of laser power for four different applied biases on a different QD than that in Fig. 2 but in the same region of the device. As the laser power is increased the electric field at the QD's position is reduced due to a space-charge effect. This behavior can be explained by charge screening at the free surface; excited electrons and holes in the depletion region are swept in opposite directions and screen the electric field. Fig. 4 shows the intensity decrease due to the electric field for the neutral exciton presented in Fig 1. This result is caused by a combination of a reduction of oscillator strength and an in-creased carrier tunneling probability. In addition, carriers excited at 850 nm must relax into the ground state. Carriers in the p-shell have an even larger tunneling probability than those in the s-shell due to the electric field, effectively reducing the pumping rate into the ground state. The tunneling of carriers out of the ground state also leads to a reduced exciton lifetime and directly contributes to an increased linewidth Γ. In addition, there may be significant spectral diffusion caused by the large fluctuation in charge at the nearby free surface. Due to the significant line broadening observed at even a modest lateral electric field, the Stark shift could not be used as the modulation in transmission spectroscopy [5].

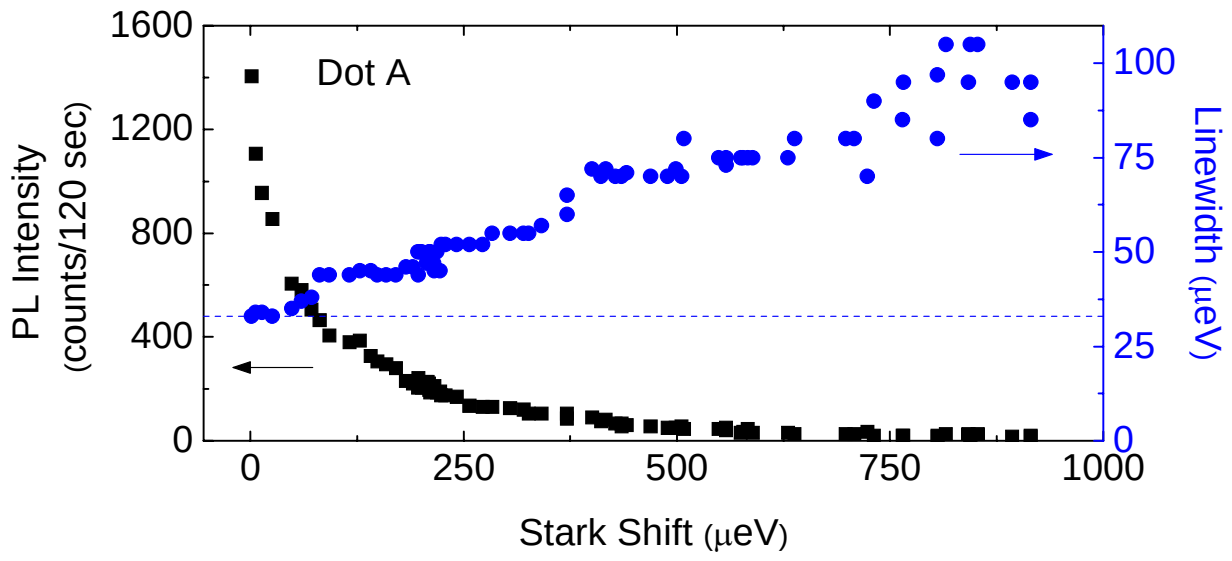

FIG. 4: PL intensity and linewidth as a function of Stark shift. With an increasing electric field, the linewidth increases and a concurrent decrease in PL intensity is observed. The dashed line corresponds to the resolution limit of the spectrometer.

Previous experiments on the FSS of self-assembled QDs have found the alignment of the linear polarization orientation to be orthogonal relative to the crystallographic fast growth direction [1-10] [3, 4, 6, 7, 15]. We have studied several QDs in the depletion region of our device, and surprisingly we find that the polarization orientation with respect to the crystal axis is not identical for each QD, similar to that reported for CdTe/ZnTe QDs [17]. Nevertheless, values for the FSS between 5 and 30 μeV at zero applied bias were observed. Fig. 5 displays the fine-structure splitting for three QDs as a function of Stark shift. The three QDs were located in the same sample in the same device and each had a PL energy near 953 nm (± 5 nm), and therefore should be very similar. As the polarization orientation is not the same for each QD, the distinction of positive or negative for the FSS values are arbitrarily chosen.

In general, two types of behavior for the FSS as a function of electric field are observed. For Dot A in Fig. 5 an intrinsic splitting of ~ 5 μeV is measured. With the application of the electric field, the FSS increases with a square root dependence on the Stark shift. At the maximum field for which there is still PL intensity, a FSS of ~ 94 μeV is measured. For Dots B and C, intrinsic FSS values of 27 and 17 μeV are measured, respectively. For these two QDs, the FSS shows evidence of an oscillatory response to the Stark shift whereas the other properties of the QDs (i.e. linewidth, intensity, energy) change monotonically with the Stark shift. In Dot B the FSS approaches zero before increasing three times. Dot C shows evidence of oscillatory behavior in which the FSS remarkably crosses zero splitting. To identify if the dipoles are rotating in response to the electric field, the polarization dependent PL was repeated at three different biases (identified by arrows in Fig. 5) for Dot C. In each case, the dipole orientation was the same.

To engineer devices suitable for entangled photon production, these results indicate that a lateral electric field is sufficient to tune the FSS to zero. However, the strong effect of carrier tunneling is a considerable drawback. Two strategies may circumvent this problem. One would be to find a QD with a small negative FSS that responds like Dot A. In this scenario a very small electric field would be needed thus tunneling is unlikely. A second approach might be to use QDs with a deeper confining potential such that the effect of carrier tunneling is reduced, although in such dots the initial FSS may be larger.

In summary, we have studied QDs in the depletion region of a Schottky gate. A lateral electric field was used to induce Stark shifts of up to 1.5 meV in the ground state neutral exciton. Additionally, a decrease in PL intensity and a linewidth broadening were observed due to the electric field. We report the manipulation of the FSS over a large range (~ 100 μeV) and most importantly through zero. We hope that these results stimulate a microscopic understanding of the electric field dependence on the exciton fine structure splitting.

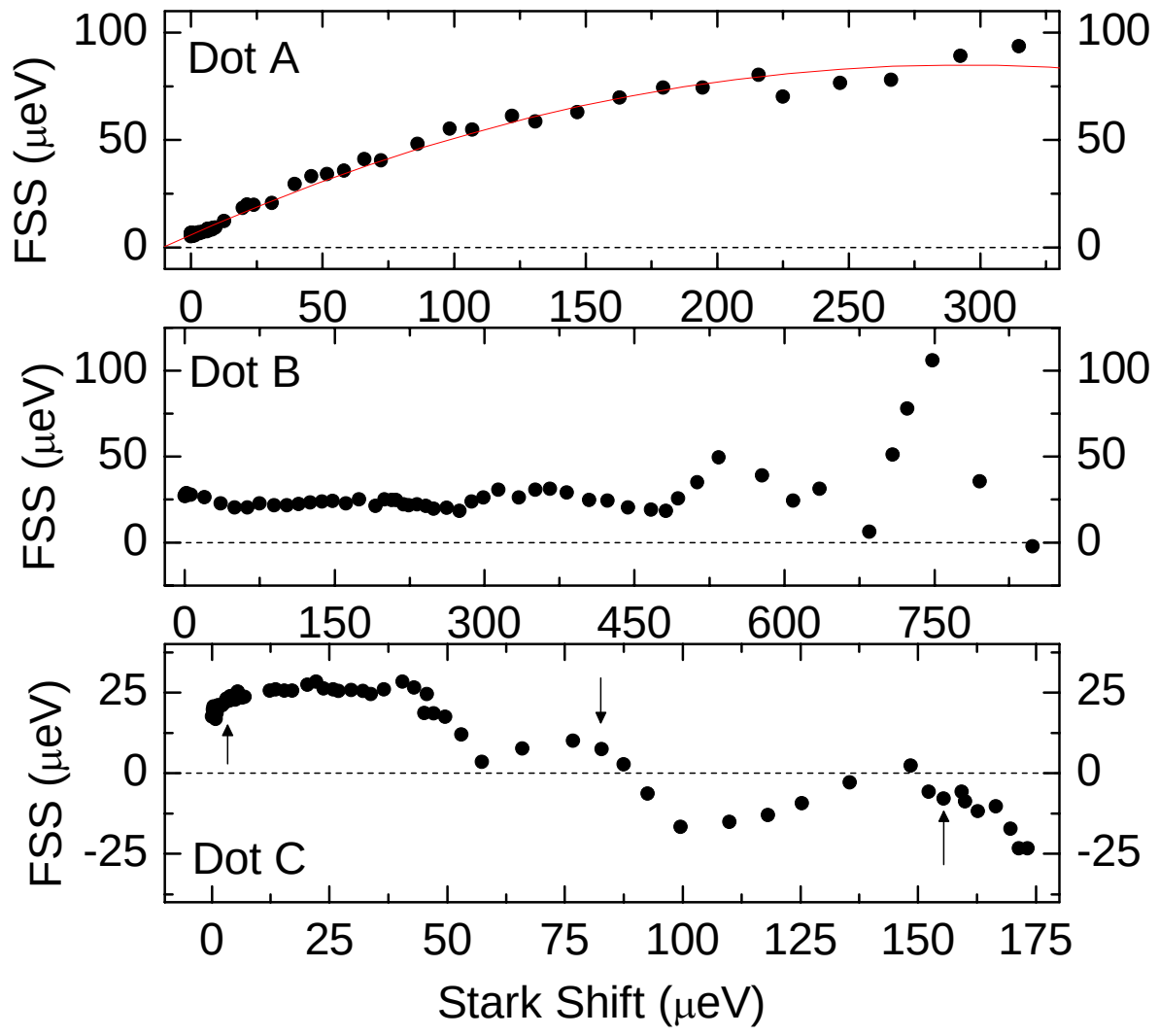

FIG. 5: The fine-structure splitting of three different QDs as a function of Stark shift. In Dot A, the splitting shows a square root dependence on the Stark shift. Dots B and C exhibit a splitting with an oscillatory response to the Stark shift. Dots A and B approach zero splitting, whereas in Dot C the splitting is tuned through zero. The arrows for Dot C correspond to the values where the dipole orientations were measured.

We acknowledge financial support for this work from SANDIE (EU), EPSRC (UK), and DAAD (Germany).